\documentclass[aps,prb,notitlepage,superscriptaddress,12pt]{revtex4-1}
\usepackage{amsfonts}
\usepackage{amsmath}
\usepackage{amssymb}
\usepackage{graphicx}
\usepackage{mathrsfs}
\usepackage{bm}
\usepackage{color}
\usepackage{float}
\usepackage[ruled,linesnumbered]{algorithm2e}
\usepackage{algorithmic} 
\usepackage{subfigure}
\renewcommand{\algorithmicrequire}{\textbf{Input:}}
 
\begin{document}

\title{Nuclear phase retrieval spectroscopy using resonant x-ray scattering}

\author{Ziyang Yuan}
\affiliation{College of Liberal Arts and Science, National University of Defense Technology, Changsha 410073, People's Republic of China}
\affiliation{Academy of Military Science, Beijing, 100097, People’s Republic of China}

\author{Hongxia Wang}
\affiliation{College of Liberal Arts and Science, National University of Defense Technology, Changsha 410073, People's Republic of China}

\author{Zhiwei Li}
\affiliation{Key Lab for Magnetism and Magnetic Materials of the Ministry of Education,  Lanzhou University, Lanzhou 730000, People's Republic of China}

\author{Tao Wang}
\affiliation{Key Lab for Magnetism and Magnetic Materials of the Ministry of Education,  Lanzhou University, Lanzhou 730000, People's Republic of China}

\author{Hui Wang}
\affiliation{Key Lab for Magnetism and Magnetic Materials of the Ministry of Education,  Lanzhou University, Lanzhou 730000, People's Republic of China}

\author{Xinchao Huang}
\affiliation{Hefei National Laboratory for Physical Sciences at Microscale and Department of Modern Physics, University of Science and Technology of China, Hefei, Anhui 230026, People’s Republic of China}

\author{Tianjun Li}
\affiliation{Hefei National Laboratory for Physical Sciences at Microscale and Department of Modern Physics, University of Science and Technology of China, Hefei, Anhui 230026, People’s Republic of China}

\author{Ziru Ma}
\affiliation{Hefei National Laboratory for Physical Sciences at Microscale and Department of Modern Physics, University of Science and Technology of China, Hefei, Anhui 230026, People’s Republic of China}
\author{Linfan Zhu}
\affiliation{Hefei National Laboratory for Physical Sciences at Microscale and Department of Modern Physics, University of Science and Technology of China, Hefei, Anhui 230026, People’s Republic of China}

\author{Wei Xu}
\affiliation{Beijing Synchrotron Radiation Facility, Institute of High Energy Physics, Chinese Academy of Sciences, Beijing, 100049, People's Republic of China}

\author{Yujun Zhang}
\affiliation{Beijing Synchrotron Radiation Facility, Institute of High Energy Physics, Chinese Academy of Sciences, Beijing, 100049, People's Republic of China}

\author{Yu Chen}
\affiliation{Beijing Synchrotron Radiation Facility, Institute of High Energy Physics, Chinese Academy of Sciences, Beijing, 100049, People's Republic of China}

\author{Ryo Masuda}
\affiliation{Faculty of Science and Technology, Hirosaki University, Bunkyo-cho, Hirosaki-shi, Aomori 036-8561 Japan}

\author{Yoshitaka Yoda}
\affiliation{Precision Spectroscopy Division, Japan Synchrotron Radiation Research Institute, Sayo, Hyogo 679-5198, Japan}

\author{Jianmin Yuan}
\email{jmyuan@gscaep.ac.cn}
\affiliation{Department of Physics, Graduate School of China Academy of Engineering Physics, 100193 Beijing, China}
\affiliation{College of Liberal Arts and Science, National University of Defense Technology, Changsha 410073, People's Republic of China}

\author{Adriana P\'alffy}
\email{adriana.palffy-buss@physik.uni-wuerzburg.de} 
\affiliation{Institute of Theoretical Physics and Astrophysics, University of W\"urzburg, Am Hubland, 97074 W\"urzburg,  Germany}
\author{Xiangjin Kong} 
\email{kongxiangjin@nudt.edu.cn}
\affiliation{College of Liberal Arts and Science, National University of Defense Technology, Changsha 410073, People's Republic of China}
%

%
\date{\today}
\maketitle

{\bf 
Light-matter interaction is exploited in spectroscopic techniques to access information about molecular, atomic or nuclear constituents of the sample of interest. While scattered light carries both amplitude and phase information of the electromagnetic field, most of the time the latter is lost in intensity measurements. However, often the phase information is paramount to reconstruct the desired information of the target, as it is well known from coherent x-ray imaging. Here we introduce a new phase retrieval algorithm which allows us to reconstruct the field phase information from two-dimensional time- and energy-resolved spectra. We apply this method to the particular case of x-ray scattering off M\"ossbauer nuclei at a synchrotron radiation source. Knowledge of the phase allows also for an excellent reconstruction of the energy spectra from experimental data, which could not be achieved with this resolution otherwise. Our approach provides an efficient novel data analysis tool which will benefit x-ray quantum optics and M\"ossbauer spectroscopy with synchrotron radiation alike. 

}

  
 X-ray scattering is a powerful tool for imaging, given the involved wavelengths which are commensurate with molecular or interatomic distances\cite{XrayPhys2011}. Resonant x-ray scattering on the other hand often involves atomic transitions of core electrons or narrow resonances of M\"ossbauer nuclei. The latter occur at x-ray wavelengths and  can be considered as ideal quantum systems with high quality factors \cite{Adams2013}. The most widely used M\"ossbauer nuclear resonance is the transition between the ground state and the first excited state of $^{57}$Fe, for which the energy resolution $\Delta E/E$, defined as the ratio of transition linewidth $\Delta E$ and resonant energy $E$, is $3\times10^{-13}$. This energy resolution is even higher for other M\"ossbauer nuclei, like $^{45}$Sc ($\Delta E/E\sim10^{-19}$) and $^{103}$Rh ($\Delta E/E\sim10^{-24}$) \cite{kalvius2012rudolf}. Such nuclear resonances can be very sensitive to their environment, providing sensitive information in various fields such as  physics, chemistry, biology or metallurgy. For instance, M\"ossbauer nuclei have been used to determine the gravitational redshift\cite{pound1959gravitational,pound1960apparent,pound1964effect}, to study  magnetically ordered materials\cite{long2013mossbauer} and to investigate the mineralogy of iron-bearing rocks and soils\cite{morris2004mineralogy}. Meanwhile, with the development of x-ray sources and detecting devices, M\"ossbauer nuclei of exceptionally narrow resonances are treated as a promising platform to implement quantum optics or coherent control in the hard x-ray regime \cite{Adams2013,heeg2021coherent}. Nuclear resonances are studied at synchrotron sources both in the energy- and in the time domain. In both cases, x-ray intensities are measured, thus loosing any information on the phase of the scattered electromagnetic radiation.

 Synchrotron radiation (SR) has high brilliance and can be well focused on micrometre-size samples. Typically, SR is used in time-resolved nuclear forward scattering\cite{hastings1991mossbauer}, which is not particularly advantageous to study materials with highly complex phases, because in the time spectra all scattering channels via spin states, valence states and crystallographic sites interfere ma{}king individual identification cumbersome. In order to obtain absorption spectra in the energy domain, a Synchrotron M\"ossbauer Source can be used\cite{chumakov1990time}, which is however at present available only at few beamlines. 
An alternative and generally accessible technique employs an additional single-line reference sample, which is mounted on a M\"ossbauer (Doppler) drive and scans the nuclear transition frequencies by varying the velocity. The detection of the prompt off-resonant pulse is suppressed exploiting the SR pulsed time structure. Delayed forward scattered photons are recorded and integrated over time as a function of the detuning of the reference sample, which recovers the energy spectrum of the sample under investigation\cite{coussement1996time,l2000experimental}. However, the accuracy of this method heavily relies on the integrated time window, which is often inflexible due to the beamline working parameters. Using a periodic time window, stroboscopic detection (SD) has been used as alternative
\cite{callens2002stroboscopic,callens2003principles,rohlsberger2012electromagnetically}. Also, very 
 recently, a  method based on a rapidly oscillating reference sample has been proposed to measure both the amplitude and the phase of the spectral response\cite{herkommer2020phase}.

Here, we demonstrate  a new method to recover both the amplitude and the phase of the electromagnetic field scattered off an unknown sample containing M\"ossbauer nuclei in a setup using a reference sample on a Doppler drive. Our nuclear phase retrieval spectroscopy (NPRS) method uses as input a full time- and energy-resolved data set provided by recording simultaneously the time of arrival and the corresponding Doppler velocity of the M\"ossbauer drive for each x-ray photon count. Instead of using time-integrated spectroscopy (TIS), the energy spectrum and the field phase are recovered using a phase retrieval formalism algorithms reminiscent of methods used to solve the phase problem in imaging\cite{XrayPhys2011,hendrickson1991determination,son2011multiwavelength}. The advantages of the NPRS methods are two-fold. First, it recovers more accurate energy spectra than TIS and SD, since it can actively extract information from all counts over the entire temporal spectrum at every velocity value. Second, it provides reliable phase information for the scattered field without introducing any physical model for the sample. We note here that two recent experiments have extracted  phase information based on physical models for the used setup \cite{evers2017,heeg2021coherent}. However, the method presented here has the advantage that it provides the phase without any assumptions on the underlying physical model. While the phase itself is helpful but not crucial for the nuclear spectra investigated here, other scattering problems may benefit from the phase retrieval algorithm introduced here.   Since it is completely model-independent and data-based, NPRS could be integrated in the data acquisition and analysis tools directly at SR beamlines.

The experimental setup and the NPRS input and output sets are illustrated in Fig.~\ref{system}. A target containing $^{57}$Fe M\"ossbauer nuclei with a resonance at 14.4 keV  is probed by a resonant but spectrally broad x-ray pulse from SR. An analyzer containing the same nuclei is mounted on a M\"ossbauer drive that provides a periodic energy detuning. Single x-ray photons are detected by fast avalanche photodiode detectors (APDs). The 8 element APDs\cite{baron2006silicon} record the counts of the photons in a given time span at each detuning.
 The delayed x-ray photons may come from three possible paths: scattered by the sample under investigation, scattered by the reference sample, or scattered by both samples. The photon scattered by both samples arrives at later time since it has been delayed twice. This input data set is used by our algorithm over several iterations to reconstruct the complex energy-dependent response of the sample.

\begin{figure}
	\includegraphics[width=1.0\textwidth]{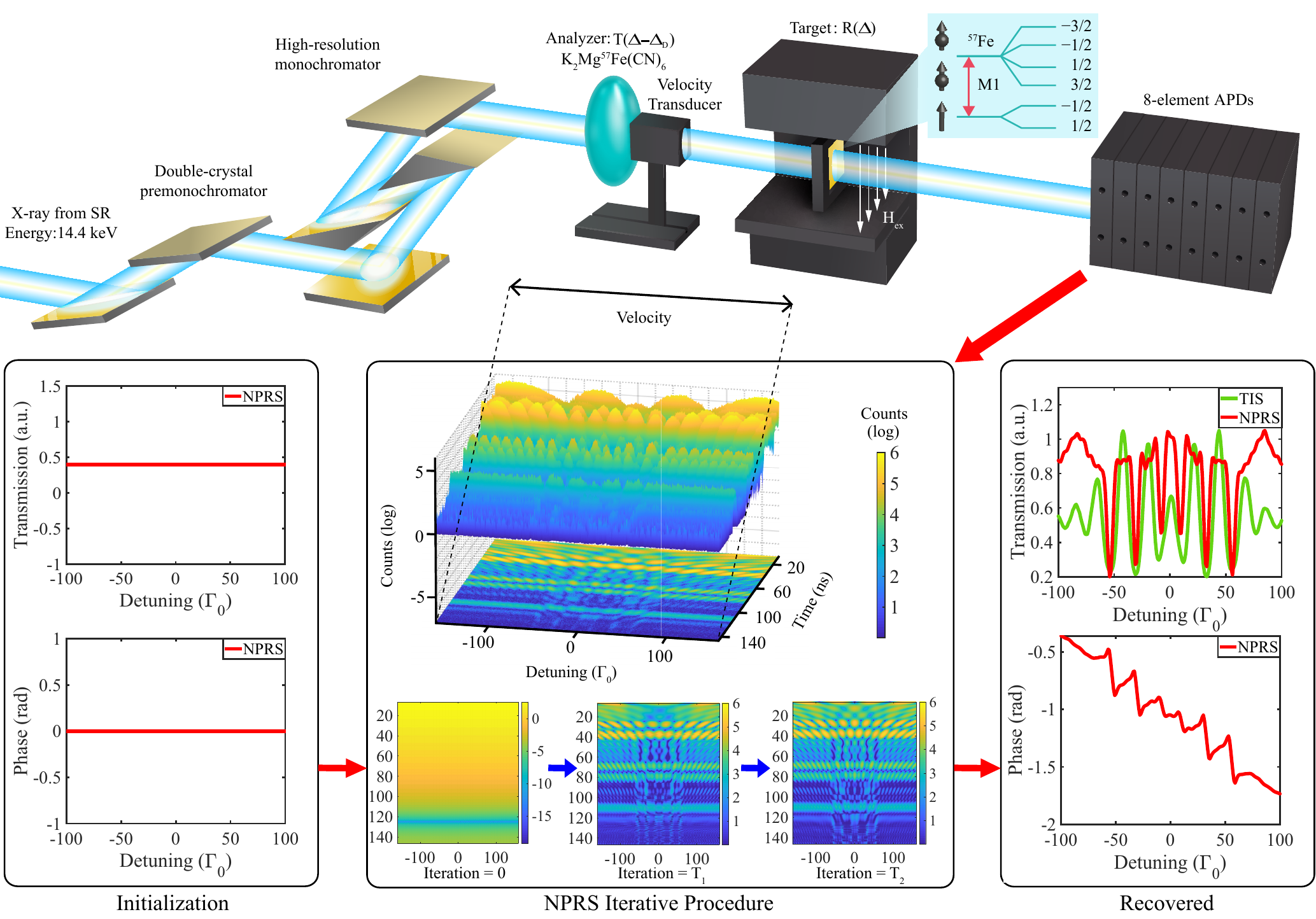}
	\caption{Schematic diagram of NPRS. Resonant monochromatized SR is scattered off a nuclear sample containing $^{57}$Fe whose response function is represented by $R(\Delta)$. The nuclear resonance width is only  4.6 neV,  such that the SR is in comparison spectrally broad. 
	 An analyzer containing the same nuclei is mounted on a M\"ossbauer drive that provides a periodic energy detuning and the transmission function is denoted by $T(\Delta-\Delta_D)$. Single x-ray photons are detected by fast avalanche photodiode detectors (APDs). The 8 element APDs record the counts of the photons in a given time span at each detuning during the experiment. Our iterative NPRS algorithm is applied to recover the M\"ossbauer spectrum of the nuclei from these registered counts.   }
	\label{system}
\end{figure} 

\textbf{Mathematical model} 

The starting point of the algorithm is the 
 time- and energy-dependent (measured) intensity $I(t,\Delta_D)$, where $\Delta_D$ is the Doppler detuning which shifts the value of the frequency seen by the analyzer foil moving with velocity $v$ as $\omega'=\omega(1+v/c)=\omega +\Delta_D$. Here, $c$ stands for the speed of light. We let $R(\Delta)$ denote the response function of the sample under investigation and with $ T(\Delta-\Delta_D)$  the analyzer foil transmission, respectively,  where $\Delta$ is the x-ray photon frequency. The mathematical formulation of the physical procedure in Fig.~\ref{system} can be modeled as
\begin{equation}\label{t1}
I(t,\Delta_D)=\left|\frac{1}{\sqrt{2\pi}} \int_{-\infty}^{\infty}  R(\Delta) T(\Delta-\Delta_D)e^{-i\Delta t}d\Delta\right|^2.
\end{equation}

During the experiment, the photon counts $I(t_k,\Delta_D^l)$ are measured at  $t_k\in\mathbf{E}=\{t_1,t_2,\cdots,t_K\}$ under different 
 Doppler detunings $\Delta_D^l\in\bm{\Omega}=\{\Delta_D^1,\Delta_D^2,\cdots,\Delta_D^L\}$. If we discretize the energy range $[-\Delta_{\textrm{max}},\Delta_{\textrm{max}}]$ by equally spaced nodes $\Delta_j$ with stepsize $\Delta_{\text{step}}$, and define the  vectors  $\mathbf{R}=(R(\Delta_0),R(\Delta_1),\cdots,R(\Delta_{n-1}))$, $\mathbf{T}_l=(T(\Delta_0-\Delta_D^l),T(\Delta_1-\Delta_D^l),\cdots,T(\Delta_{n-1}-\Delta_D^l))$, the discretized model of \eqref{t1} can be  formulated as 
\begin{eqnarray}
I(t_k,\Delta_D^l)=\left|\mathbf{F}_{:,k}^{\text{H}}(\mathbf{R}\odot\mathbf{T}_{l})\right|^{2}+\varepsilon, t_k\in\mathbf{E},\Delta_D^l\in\bm{\Omega},\label{discrete model}
\end{eqnarray}
where $\mathbf{F}_{:,k}=\frac{\Delta_{\textrm{step}}}{2\pi}(1,e^{i1\Delta_{\textrm{step}}\cdot t_k},\cdots,e^{i(n-1)\Delta_{\textrm{step}}\cdot t_k}),$ $\odot$ is the Hadamard product, $(\cdot)^{\text{H}}$ represents the conjugate transpose, and $\varepsilon$ is the error caused by  noise and discretization (see Supplemental Material for details). 

We abbreviate $I(t_k,\Delta_D^l)$ to $I(k,l)$ which denotes the $(k,l)$th element of the matrix $\mathbf{I}$, and then estimate $\mathbf{R}$ from $\mathbf{I}$ by Bayesian method, which maximizes the $\text{log}$ of the posterior conditional probability $P(\mathbf{R}\mid\bm{\mathrm{I}})$. According to the Bayes' rule, we have $P(\mathbf{R}\mid \bm{\mathrm{I}})\propto \overset{(a)}{P(\bm{\mathrm{I}}\mid \mathbf{R})}\overset{(b)}{P(\mathbf{R})},$
where (a) is the likelihood which usually satisfies the Poisson distribution, and (b) allows to introduce the prior knowledge of $\mathbf{R}$. In practice, the physical model about $\mathbf{R}$ is not always available. Here, we omit (b) and consider a more general case without using any prior information of $\mathbf{R}$. The optimization model to recover $\mathbf{R}$ from $\bm{\mathrm{I}}$ is established as 
\begin{eqnarray}\label{optimization}
\mathop{\textrm{minimize}}_{\mathbf{R}\in\mathbb{C}^n}\ell(\mathbf{R})=\sum_{k=1}^{K}\sum_{l=1}^{L}\left(\left|\mathbf{F}_{:,k}^{\text{H}}(\mathbf{R}\odot\mathbf{T}_{l})\right|^{2}-I(k,l)\text{log}\left(\left|\mathbf{F}_{:,k}^{\text{H}}(\mathbf{R}\odot\mathbf{T}_{l})\right|^{2}\right)\right).
	\end{eqnarray}  
Eq.~\ref{optimization} is a challenging non-convex optimization problem since $\ell(\mathbf{R})$ is nonconvex and thus many local minimums may exist\cite{shechtman2015phase,chen2017solving}. In our manuscript, a gradient based algorithm combined with the momentum restart and adaptive reweighed modules is established, which could recover both the amplitude and the phase of response function and be treated as NPRS method (see Methods for details).

\textbf{Numerical simulation}

In order to test the potential and the broad applicability of the NPRS method described above, we apply it first to simulated data including three scenarios of resonant nuclear x-ray scattering which have been demonstrated experimentally so far: (i) normal incidence targets containing 
 $\alpha$-Fe with different magnetization orientations, (ii) the recently demonstrated setup in Ref.~\cite{evers2017} employing a forward-scattering target combined with fast mechanical motion which reshuffle spectral components of the x-ray pulse, and (iii) a thin-film cavity setup similar to the one in Ref.~\cite{rohlsberger2010collective}. The nuclear targets for (i) and (ii) above are in free space and the transmission spectra are considered as input for the phase retrieval formalism. The simulations for all scenarios were obtained using a numerical implementation of the Parratt formalism that we have benchmarked with the software packages CONUSS \cite{sturhahn2000conuss} and Pynuss. The simulated response function $R(\Delta)$ is used  in Eq.~\eqref{t1} to generate the input data for the phase retrieval algorithm.

\begin{figure}
	\includegraphics[width=1.0\textwidth]{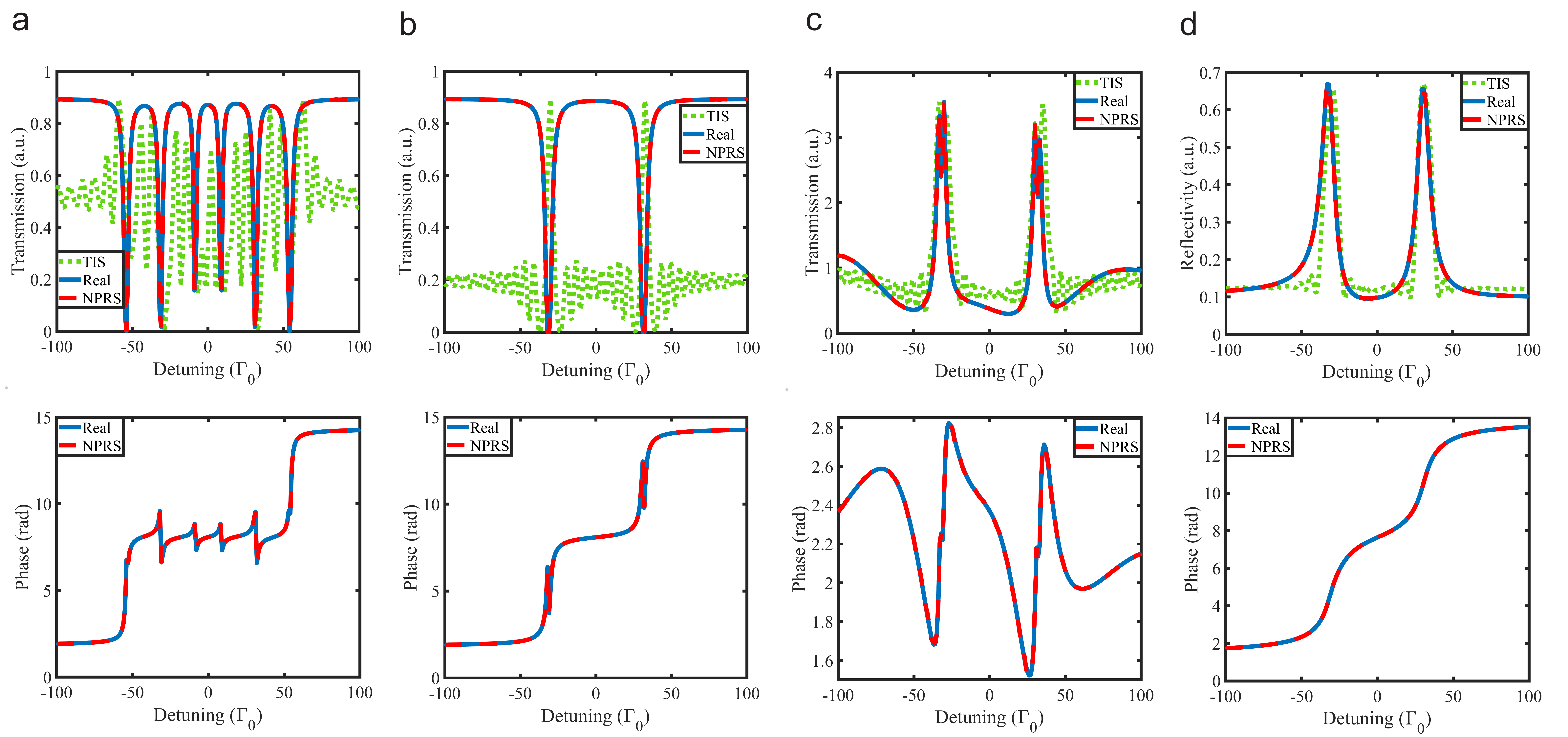}
	\caption{Simulation results for transmission and phase as a function of detuning. We consider  an $\alpha$-Fe sample of $2.3\, \mu m$ with \textbf{a} random magnetization and  \textbf{b} an aligned magnetization. \textbf{c} illustrates the case when a fast mechanical motion with the displacement $0.43$\,\r{A} is applied on the nuclear target. \textbf{d} presents the results for a thin-film cavity with the structure  Ta (2 nm)/B (17 nm)/$^{57}\mathrm{Fe}$(3 nm)/B (7 nm)/Ta (9 nm) at the incidence angle  $3.5\, \text{mrad}$. Here $^{57}\mathrm{Fe}$ is considered as the resonant M\"ossbauer nucleus for all the targets. For each case, the simulated and NPRS-recovered intensities and phases are presented in the top and lower rows of the figure,  while the top row also shows the recovered spectrum by the TIS method. }
	\label{num}
\end{figure}

For the simulation, we consider the single-resonance $\mathrm{K}_2\mathrm{Mg}^{57}\mathrm{Fe}(\mathrm{CN})_6$ analyzer with the effective thickness $d\approx1\,\mu m$ that we have later also used in our experiment. The intensity which would be recorded in experiment is calculated by Eq.~\ref{t1}. In order to simulate the experimental conditions at the SR beamline, the time range is set from 3 ns to 165 ns. The signals before 3 ns are excluded since the off-resonant component of the incident x-ray pulse is very strong. The upper limit 165 ns corresponds to the bunch separation of the E mode at the nuclear resonant scattering beamline BL09XU of SPring-8 in Japan. The developed NPRS method is applied on the truncated data. NPRS delivers the energy transmission spectra and the phase as a function of the detuning, which are then compared with the simulated data ones in Fig.~\ref{num}, demonstrating an excellent agreement. NPRS recovers both the amplitude and phase of the target spectra with high accuracy.
 
For scenario (i), we consider  the situation that a  linearly polarized x-ray pulse irradiates the target. When the target is randomly magnetized, six transitions of different energies occur in the spectrum, as shown in Fig.~\ref{num} \textbf{a}. By applying a weak external magnetic field, the magnetization orientation of the target can be aligned. When the magnetization orientation is perpendicular to the propagation and polarized directions of input x-ray pulses, two transitions are driven, as shown in Fig.~\ref{num}\textbf{b}. These are two typical cases in M\"ossbauer spectroscopy. The results for case (ii) are presented in Figs.~\ref{num}\textbf{c}. In Fig.~\ref{num}\textbf{d}, we consider a  thin-film cavity containing a $^{57}$Fe layer, with a structure similar to  the ones used in Refs.\cite{PhysRevLett.111.073601, heeg2015tunable}.

Both the intensity and the phase recovered by NPRS are in excellent agreement with the real values for all the four cases considered here. 
For comparison, we also use the well-establish TIS method 
 which recovers the intensity spectrum by integrating the counts over late times. The TIS method was widely used in recent experiments\cite{rohlsberger2010collective,heeg2013vacuum,haber2016,evers2017,haber2017rabi}.  The quality of the recovered spectra depends on the integration window and the thickness of the analyzer. In order to obtain the best possible results, we optimize the integration range  for each example separately, however using the same analyzer parameters. We find that TIS does not work well for the cases presented in Fig.~\ref{num} \textbf{a} and \textbf{b} and it recovers the spectrum of the cases presented in Fig.~\ref{num} \textbf{c} and \textbf{d} with less good quality than NPRS. We note that the quality of the recovered spectrum by the TIS method could be improved by optimizing the thickness of the analyzer. However, for consistency here we stick to the analyzer thickness  that was actually used in our experiments.

Compared with the TIS method, the simulation results present the key advantages of the NPRS method: it uses all the counts within the measured time range, not only the low count rates in later time range, but also the much higher counts rates from earlier time. Second, it does not depend on the choice of the integration window which makes the NPRS method more objective. Obviously, NPRS remains sensitive on the quality and range of the data set. Third, it is much more robust against the thickness of the analyzer that makes the NPRS method more feasible for a real experiment. More importantly, it provides the phase information of the nuclear target, which is inaccessible in TIS method. This allows us to reconstruct the 2D intensity, which provides a way to further test the recovered spectrum.

\textbf{Experimental results}

\begin{figure}
	\includegraphics[width=1.0\textwidth]{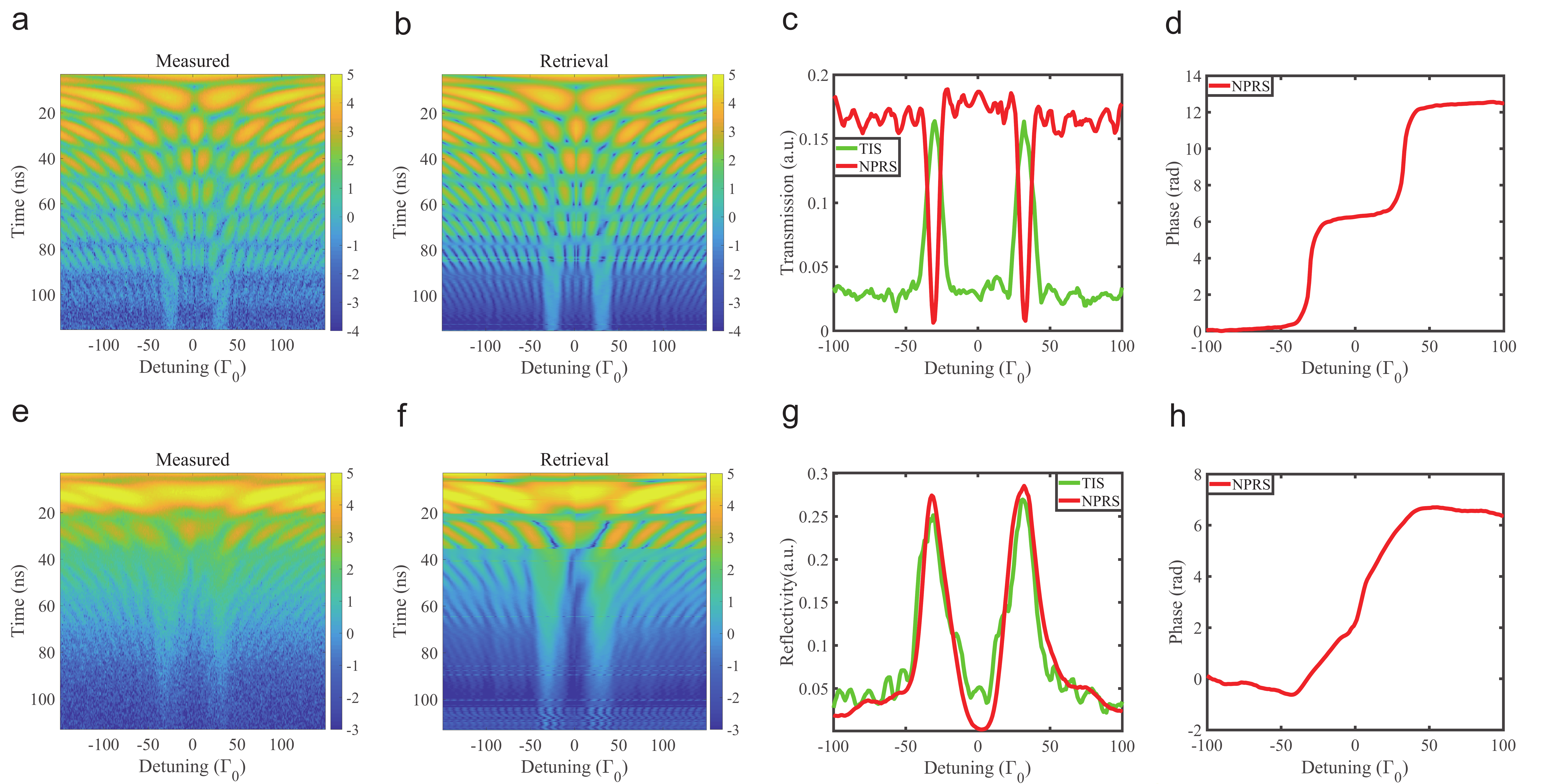}
	\caption{Experimental and reconstructed results. \textbf{a}-\textbf{d} present the results for the $\alpha$-Fe target with aligned magnetization. \textbf{e}-\textbf{h} show the case of the thin-film cavity with one nuclear layer containing $^{57}\mathrm{Fe}$. The measured and NPRS reconstructed 2D intensity spectra are shown in \textbf{a} and \textbf{b} (\textbf{e} and \textbf{f}), respectively. The recovered spectra by NPRS and TIS methods are presented in \textbf{c} (\textbf{g}). The NPRS-recovered phase is shown in \textbf{d} (\textbf{h}).}
	\label{exp}
\end{figure}

Experiments were performed for the scenarios (i) and (iii) discussed above 
 at the nuclear resonant scattering beamline BL09XU of SPring-8 in Japan. To measure the 2D intensity data, an event-based data acquisition system was used to record temporal (time of arrival) and spectral (Doppler drive velocity) information for each signal photon separately. 
A single-line  $\mathrm{K}_2\mathrm{Mg}^{57}\mathrm{Fe}(\mathrm{CN})_6$  analyzer was used, whose time response $T(\Delta-\Delta_D)$ was determined from a measured time spectrum (see Methods). 
The effective thickness of the analyzer was deduced to be approx. $d\approx1\,\mu m$. A small external magnetic field was used to align the magnetization of the targets in the two considered scenarios. The experimental results are shown in Fig.~\ref{exp}. The measured 2D data of a $\alpha$-Fe sample and a thin-film cavity with one nuclear layer are presented is Fig~.\ref{exp} \textbf{a} and \textbf{e}, respectively. Correspondingly, the reconstructed 2D data by NPRS method are shown in Fig.~\ref{exp} \textbf{b} and \textbf{f}, and are in good agreement with the experimental data. The recovered spectra by NPRS method are shown in Fig.~\ref{exp} \textbf{c} and \textbf{g}, in which the results obtained by TIS method are presented as a reference. As already presented in the simulation results of Fig.~\ref{num} \textbf{b} and \textbf{d}, the two dips presented in the spectrum of the $\alpha$-Fe sample appear as peaks in the recovered spectrum by the TIS method for the thickness of the analyzer considered in this manuscript. The NPRS method works well and the two dips are clearly seen. For the cavity sample, both NPRS and TIS methods recover the two peaks which stand for the two nuclear transitions. The phase spectra recovered by NPRS method are shown in Fig.~\ref{exp} \textbf{d} and \textbf{h}, and are consistent with the simulation results presented in the previous Section. This demonstrates the power of the NPRS method to retrieve high-quality energy spectra and phase information directly from experimental data. 

\textbf{Discussion}

The NPRS method has proven its strength for retrieval of reliable phase  and  energy spectra in resonant nuclear x-ray scattering.  Its particular advantage is that accurate results can be achieved without any physical model input for the sample response function. Knowledge of the phase can be important either for the more accurate fit of the 2D data sets, or for specific applications which involve phases via mechanical motion\cite{evers2017,heeg2021coherent}. However, the true potential of the method might extend beyond the field of resonant nuclear scattering to the more general x-ray scattering. The phase problem in diffraction can be tackled making use of resonant scattering for instance in the example of multi-wavelength anomalous diffraction\cite{XrayPhys2011,hendrickson1991determination,son2011multiwavelength}. By adapting our method for two-dimensional energy- and angle-spectra, we should be able to obtain reliable phase information in a model-independent fashion. As such, the data-based NPRS method could be in the future directly integrated in the data acquisition and analysis tools  at SR x-ray imaging beamlines.

\clearpage
%
 
%

\subsection*{METHODS}
\textbf{Algorithms.} The core of NPRS method is the vanilla gradient descent (GD) algorithm, which is one of the 
 most popular optimization algorithms for the non-convex problems \cite{2017Implicit} and by far the most widespread used method in deep learning \cite{2015Deep}. By choosing a proper initialization $\mathbf{R}^{(0)}$, the $m$th iteration takes the form 
$$
\mathbf{R}^{(m)}=\mathbf{R}^{(m-1)}-\mu\sum_{k=1}^{K}\sum_{l=1}^{L}\left(1-\frac{I(k,l)}{\left|\mathbf{F}_{:,k}^{\text{H}}(\mathbf{R}^{(m-1)}\odot\mathbf{T}_{l})\right|^{2}}\right)(\mathbf{F}_{:,k}\odot\overline{\mathbf{T}_l})(\mathbf{F}_{:,k}\odot\overline{\mathbf{T}_l})^{\text{H}}\mathbf{R}^{(m-1)},
$$
where $\mu$ is the step size. Furthermore, momentum with flexible parameter restart \cite{2012Adaptive} and adaptive reweighed \cite{2018Solving} techniques are utilized to accelerate the convergence and improve the performance.  The iterations will not stop until  $\ell(\mathbf{R}^{(m)})$ is below some given error bound (more details are shown in Supplemental Material). 

\begin{figure}
    \includegraphics[width=0.75\textwidth]{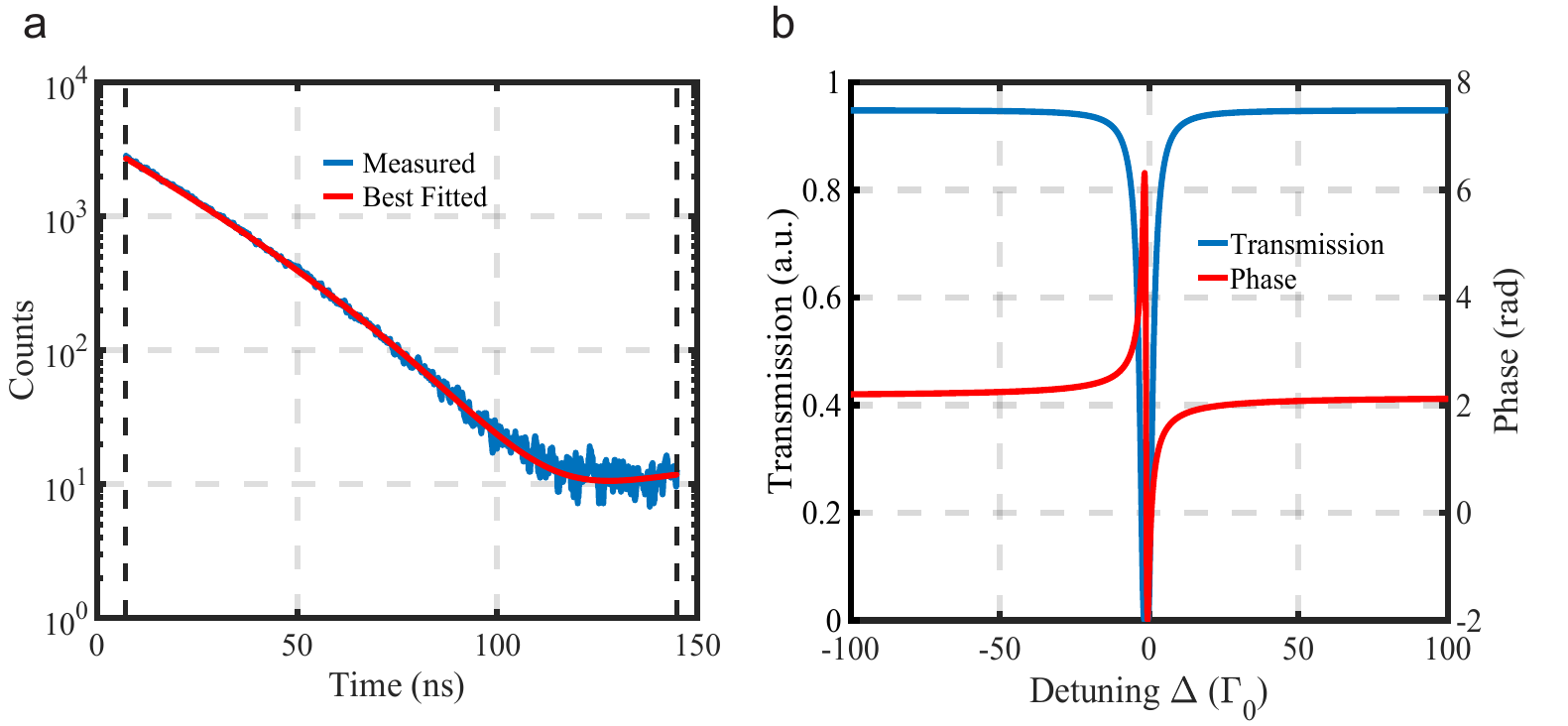}
    \caption{\label{analyzer} \textbf{a}, Fit of the time spectrum of the analyzer. \textbf{b}, The transmission and phase of the fitted response function when $\Delta_D=0.$ }
    \label{ana} 
\end{figure}

\textbf{Analyzer response function.} 
In order to obtain the response function of the analyzer, we fit a measured time spectrum (see Fig.~\ref{ana}). The isomer shift of the analyzer had been previously measured to be  -0.1 mm/s relative to $\alpha$-$\mathrm{Fe}$. The obtained fitted response function is $T(\Delta-\Delta_D)=(-0.54 + 0.81i)e^{-\frac{4.16\Gamma_0i}{\Delta-\Delta_D+1.03\Gamma_0+0.5\Gamma_0i}}$, where $\Gamma_0$ is the spontaneous decay rate of $^{57}\mathrm{Fe}$. As an example, the transmission and phase of the analyzer for the case when $\Delta_D=0$ is shown in Fig.~\ref{ana}.

\textbf{Sample preparation and measurements.} 
The thin-film cavity used in the experiment was deposited on surface polished Si substrates (dimensions $20\times20\,\text{mm}^2$) by DC magnetron sputtering technique. A base pressure of $5\times10^{-5}$ Pa was achieved before the cavity layer deposition and pure Argon gas of 0.3 Pa was used as the sputtering working gas for all the deposited layers. For the $^{57}\mathrm{Fe}$ layer, an enriched $^{57}\mathrm{Fe}\left(\geq95\%\right)$ ring slightly bigger than the erosion race-track was placed on the natural Iron target. All other layers were deposited by pure element targets.

The experiment was performed at the Nuclear Resonant Scattering Beamline BL09XU of SPring-8 in a bunch mode providing a separation of 165.2 ns between the x-ray pulses (mode-E) and a 2/29-filling bunch train which is blocked by the timing electronics. The pulse duration is determined by the length of the electron bunches in the storage ring and amounts to about 40 ps.

\subsection*{Acknowledgements}
Z.Y. and H.X.W. acknowledge support by National Key Research and Development Program under Grant No. 2020YFA0713504, National Natural Science Foundation of China under Grant No. 61977065 and 173 Program of China under Grant No.2020-JCJQ-ZD-029. X.K. acknowledges support by the National Natural Science Foundation of China (NSFC) under Grant No. 11904404. A.P. gratefully acknowledges support from the Heisenberg Program of the Deutsche Forschungsgemeinschaft (DFG). J.Y. acknowledges the support by National Key Research and Development Program of China under Grant No. 2019YFA0307700. Z.L. acknowledges support by NSFC under Grant No. 11704167. L.Z. acknowledges support by NSFC under Grant No. U1932207. W. X and Y.J.Z. are supported by the high-energy photon source (HEPS), a major national science and technology infrastructure. We acknowledge the support by the nuclear resonant scattering beamline BL09XU of SPring-8 in Japan with Proposal No. 2020A1101. We acknowledge the allocation of beamtime by beamline 1W1A of Beijing Synchrotron Radiation Facility (BSRF), as well as the support from Dr. Jialu Wu, Prof. Jing Ma, Prof. Yuanhua Lin and Prof. Cewen Nan from Tsinghua University, for the sample reflectivity measurements.

\subsection*{Author contributions}
X.K., J.Y. and A.P. proposed this project and devised the experimental concept. Z.Y. and H.X.W. developed the mathematical model and the algorithms. R.M. and Y.Y set up the beamline instrumentation and data acquisition systems and operated the beamline. Z.L., T.W. and H.W. fabricated the samples. Z.Y. and X.K. performed the numerical simulations and the experimental data analysis. X.K., Z.Y., H.X.W. and A.P. wrote the manuscript. All authors participated in performing the experiment, discussing the experimental results and editing the manuscript.



\subsection{Supplementary information}

\section{Error Analysis of the discrete model}
We recall that the photon counts  $\{I(t,\Delta_D)\}_{t\in\mathbf{E},\Delta_D\in\bm{\Omega}}$ are measured at a series of discrete time points $\mathbf{E}=\{t_1,t_2,\cdots,t_K\}$, where $\bm{\Omega}=\{\Delta_D^1,\Delta_D^2,\cdots,\Delta_D^L\}$. For each $\Delta_D$, the corresponding function of the analyzer $T(\Delta-\Delta_D)$ is known in advance. The continuous model 
\begin{eqnarray}\label{y1}
I(t,\Delta_D)=\left|\frac{1}{\sqrt{2\pi}} \int_{-\infty}^{+\infty}  R(\Delta) T(\Delta-\Delta_D)e^{-i\Delta t}d\Delta\right|^2
\end{eqnarray}
can be discretized to
\begin{eqnarray}\label{dis}
I(t_k,\Delta_D^l)=\left|\mathbf{F}_{:,k}^{\text{H}}(\mathbf{R}\odot\mathbf{T}_{l})\right|^{2}+\varepsilon, t_k\in\mathbf{E},\Delta_D^l\in\bm{\Omega},
\end{eqnarray}
with $\mathbf{F}_{:,k}=\frac{\Delta_{\textrm{step}}}{2\pi}(e^{i0\Delta_{\textrm{step}}\cdot t_k},e^{i1\Delta_{\textrm{step}}\cdot t_k},\cdots,e^{i(n-1)\Delta_{\textrm{step}}\cdot t_k})$. The error $\varepsilon$ in \eqref{dis} comes from two sources, aliasing and truncation. 

The aliasing part results from sampling the continuous variables. Note that any discrete signal can be represented as a sum of delta functions. For $R(\Delta)$, we associate any sample $R(h\Delta_{\text{step}})$ with $R(\Delta)\delta(\Delta-h\Delta_{\text{step}})$ located at $\Delta=h\Delta_{\text{step}}$. As a result, a uniform sampling of $R(\Delta)$ thus corresponds to the weighted sum of delta functions below
\begin{eqnarray}
R_d(\Delta):=\sum_{h=-\infty}^{+\infty}R(\Delta)\delta(\Delta-h\Delta_{\text{step}}).\label{ss1}
\end{eqnarray}
Similarly, a uniform sampling of $T(\Delta)$ is formulated as
\begin{eqnarray}
T_d(\Delta):=\sum_{h=-\infty}^{+\infty}T(\Delta)\delta(\Delta-h\Delta_{\text{step}}).\label{ss2}
\end{eqnarray}
Next, we define
\begin{eqnarray*}I_d(t,\Delta_D)&:=&\left|\frac{\Delta_{\text{step}}}{\sqrt{2\pi}}\sum_{h=-\infty}^{+\infty}R(\Delta)T(\Delta-\Delta_D)e^{-ih\Delta_{\text{step}}t}\right|^2.
\end{eqnarray*}
By using \eqref{ss1} and \eqref{ss2}, we have 
\begin{eqnarray*}
	I_d(t,\Delta_D)&=&\left|\frac{\Delta_{\text{step}}}{\sqrt{2\pi}} \sum_{h=-\infty}^{+\infty} R(\Delta) T(\Delta-\Delta_D) \int_{-\infty}^{+\infty}\delta(\Delta-h\Delta_{\mathrm{step}})e^{-i\Delta t}d\Delta\right|^2	
	\\&=&\left|\frac{\Delta_{\text{step}}}{\sqrt{2\pi}} \int_{-\infty}^{+\infty}  R_d(\Delta) T_d(\Delta-\Delta_D)e^{-i\Delta t}d\Delta\right|^2,
\end{eqnarray*}
which is the intensity of the Fourier transform of
$R_d(\Delta)T_d(\Delta-\Delta_{D})$. To further study the effect caused by discretization, the relationship between $I_d(t,\Delta_D)$ and $I(t,\Delta_D)$ holds below
\begin{eqnarray}\label{t1}
\begin{split}
I_d(t,\Delta_D)&=\left|\frac{\Delta_{\text{step}}}{\sqrt{2\pi}} \int_{-\infty}^{+\infty}  R_d(\Delta) T_d(\Delta-\Delta_D)e^{-i\Delta t}d\Delta\right|^2\\
&=\left|\frac{\Delta_{\text{step}}}{\sqrt{2\pi}} \int_{-\infty}^{+\infty}  \sum_{h=-\infty}^{+\infty}R(\Delta)T(\Delta-\Delta_D)\delta(\Delta-h\Delta_{\text{step}})e^{-i\Delta t}d\Delta\right|^2\\
&\overset{(1)}{=}\left|\frac{1}{\sqrt{2\pi}}\sum_{p=-\infty}^{+\infty}\int_{-\infty}^{+\infty}R(\Delta)T(\Delta-\Delta_D)e^{-i\Delta(t-\frac{p}{\Delta_{\text{step}}})}d\Delta\right|^2\\
&=\left|\overset{(a)}{\frac{1}{\sqrt{2\pi}}\int_{-\infty}^{+\infty}R(\Delta)T(\Delta-\Delta_D)e^{-i\Delta t}d\Delta}\right.\\
&+\left.\overset{(b)}{\frac{1}{\sqrt{2\pi}}\sum_{p\neq 0}\int_{-\infty}^{+\infty}R(\Delta)T(\Delta-\Delta_D)e^{-i\Delta(t-\frac{p}{\Delta_{\text{step}}})}d\Delta}\right|^2,
\end{split}
\end{eqnarray}
where (1) is deduced by the equality
$$\sum_{h=-\infty}^{+\infty} \delta\left(\Delta-h\Delta_{\mathrm{step}}\right)=\frac{1}{\Delta_{\mathrm{step}}}\sum_{p=-\infty}^{+\infty} e^{i\Delta\frac{p}{\Delta_{\mathrm{step}}}},$$ $p$ and $h$ are both integers. When $\Delta_D$ is given, \eqref{t1} shows that $I_d(t,\Delta_D)$ is a $\frac{1}{\Delta_{\text{step}}}$ periodical signal which is equal to the intensity of all the summations of (a) shifted by $\frac{p}{\Delta_{\text{step}}}$. Thus, (b) is the source of aliasing when discretizing $R(\Delta)$ by a uniform step $\Delta_{\mathrm{step}}$. For (b), if $R(\Delta)T(\Delta-\Delta_D)$ is a band-limited signal with cutoff time $t_{\text{c}}$, then the influence made by the term (b) can be diminished when $\Delta_{\text{step}}\leq\frac{1}{t_{\text{c}}}$. Except for the induced error by (b), the energy resolution of the recovered spectrum and the computational complexity are also taken into account and $\Delta_{\text{step}}=1\Gamma_0$ is adopted in our calculation.

Next, the error caused by the truncation is attribute to the limitation of digital devices where $\Delta$ is in the finite range $[-\Delta_{\text{max}},\Delta_{\text{max}}]$. We divide $I_d(t,\Delta_D)$ into two parts:
\begin{eqnarray}\label{t2}
\begin{split}
I_d(t,\Delta_D)&=\left|\overset{(c)}{\frac{\Delta_{\text{step}}}{\sqrt{2\pi}}\sum_{h=0}^{n-1}R(-\Delta_{\text{max}}+h\Delta_{\text{step}})T(-\Delta_{\text{max}}+h\Delta_{\text{step}}-\Delta_D)e^{-ih\Delta_{\text{step}}t}}\right.\\
&+\left.\overset{(d)}{\frac{\Delta_{\text{step}}}{\sqrt{2\pi}}\sum_{h\in(-\infty,0)\cup[n,+\infty)}R(-\Delta_{\text{max}}+h\Delta_{\text{step}})T(-\Delta_{\text{max}}+h\Delta_{\text{step}}-\Delta_D)e^{-ih\Delta_{\text{step}}t}}\right|^2,
\end{split}
\end{eqnarray}
where (d) is usually truncated. So when $\Delta_{\text{step}}$ is given, increasing $\Delta_{\text{max}}$ can diminish the truncated error. But $\Delta_{\text{max}}$ cannot be arbitrary large, which must satisfy $\Delta_{\text{max}}=\frac{\pi}{t_{\text{step}}}$ with $t_{\text{step}}$ being the smallest sampling interval.

Above all, when the influence caused by (b) and (d) get controlled, specifically let $t_{\text{step}}$ be small and $\Delta_{\text{step}}\leq \frac{1}{t_c},$ then $I(t_k,\Delta_D^l)\approx\left|\mathbf{F}_{:,k}^{\text{H}}(\mathbf{R}\odot\mathbf{T}_{l})\right|^{2}$ holds with high accuracy.

\section{Analysis of the Solution}
In this section, the solution of the optimization problem \eqref{tt2} will be discussed, which infers $\mathbf{R}$ from counts $\mathbf{I}$, namely, 
\begin{eqnarray}
\begin{split}
\text{Find}&~\mathbf{R}\in\mathbb{C}^n\\
\text{s.t.}~
I(k,l)&=\left|\mathbf{F}_{:,k}^{\text{H}}(\mathbf{R}\odot\mathbf{T}_{l})\right|^{2}+\varepsilon.\label{tt2}
\end{split}
\end{eqnarray}
In general, the solution of \eqref{tt2} is not unique. One reason is that the global phase of $\mathbf{R}$ cannot be determined. For instance, if $\mathbf{R}_0$ is a solution of \eqref{tt2}, then $\mathbf{R}_0$ with a global phase $\theta\in[0,2\pi)$, $\mathbf{R}_0e^{i\theta}$ also satisfies the constraints in \eqref{tt2}. Note that the global phase $\theta$ is not under consideration in the experiment. Thus, when it comes to the unique solution of \eqref{tt2}, we treat the solutions differing by a global phase as the same.  

In Refs.\cite{bendory2017non,alaifari2021uniqueness,grohs2019stable}, theoretical results proved that the solution of \eqref{tt2} can be unique up to the global phase under some suitable assumptions. For example, the number of sampling points $K\geq n$, and the union of the windows $W_l$ of $|\mathbf{T}_l|^2,l=1,\cdots,L$ covers $[-\Delta_{\text{max}},\Delta_{\text{max}}]$. But these assumptions above usually fail in practice. For instance, in order to relieve the error $\varepsilon$ caused by the discretization, we must set $\Delta_{\textrm{step}}$ and $t_{\textrm{step}}$ properly. This usually leads to $K$ being less than $n$. At the same time, $\underset{l=1}{\overset{K}{\cup}}W_l$ usually cannot cover $[-\Delta_{\text{max}},\Delta_{\text{max}}].$ Thus, recovering $\mathbf{R}$ uniquely from \eqref{tt2} is nearly impossible based on these under-determined conditions discussed above. Fortunately, in the practice what we concern is a portion of $\mathbf{R}$ which is covered by $\underset{l=1}{\overset{K}{\cup}}W_l$.  For example, the region of interest $\mathbf{O}$ in $\mathbf{R}$ is $[-100\Gamma_0,100\Gamma_0]$ and $\underset{l=1}{\overset{K}{\cup}}W_l$ covers $[-170\Gamma_0, 170\Gamma_0]$ in the main text. This may guarantee $\mathbf{R}$ to be unique in the region $\mathbf{O}$, and the conjecture is verified by many numerical simulations which will be given in the last part of Section \ref{section3}.

\section{Nuclear Phase Retrieval Spectroscopy}\label{section3}

The Nuclear Phase Retrieval Spectroscopy (NPRS) method
is used to solve the optimization problem 
\begin{eqnarray}\label{optimization}
\mathop{\textrm{minimize}}_{\mathbf{R}\in\mathbb{C}^n}\ell(\mathbf{R})=\sum_{k=1}^{K}\sum_{l=1}^{L}\left(\left|\mathbf{F}_{:,k}^{\text{H}}(\mathbf{R}\odot\mathbf{T}_{l})\right|^{2}-I(k,l)\text{log}\left(\left|\mathbf{F}_{:,k}^{\text{H}}(\mathbf{R}\odot\mathbf{T}_{l})\right|^{2}\right)\right).
\end{eqnarray}  
The sketch of NPRS is shown in Algorithm \ref{NPRS}. Next, details about four aspects of the NPRS method will be analyzed.

\begin{algorithm}
	\caption{Nuclear Phase Retrieval Spectroscopy (NPRS)}
	\label{NPRS}
	\begin{algorithmic} 
		\REQUIRE 
		\STATE $\bm{\Omega}$: the set of shifting detuning $\bm{\Omega}=\{\Delta_{D}^1,\Delta_D^2,\cdots,\Delta_D^{L}\}$
		\STATE	$\mathbf{E}$: the set of discrete time set $\mathbf{E}=\{t_1,t_2,\cdots,t_K\}$
		\STATE	$\mathbf{T}_l\in\mathbb{C}^{n},l=1,\cdots,L$: the analyzer with $L$ different Doppler detunings
		\STATE	$\mathbf{I}\in\mathbb{R}^{K\times L}$: the counts of photons
		\STATE	$\mu$:~~the step size
		\STATE	$\varepsilon$:~~the error bound
		\ENSURE 
		\STATE	$\hat{\mathbf{R}}$: an estimation of $\mathbf{R}$
		\renewcommand{\algorithmicrequire}{\textbf{Initialization:}}
		\REQUIRE
		\STATE $\mathbf{R}^{(0)}\in\mathbb{C}^n$ is the initialization
		\STATE $\tilde{\mathbf{R}}^{(0)}=\mathbf{R}^{(0)}$
		\STATE $t=1$
		\renewcommand{\algorithmicrequire}{\textbf{General~Step}}
		\REQUIRE($m=1,2,\cdots$):
		\STATE $
		\tilde{\mathbf{R}}^{(m)}=\mathbf{R}^{(m-1)}-\mu \nabla \ell_{\omega}\left(\mathbf{R}^{(m-1)}\right)
		$
		\STATE $\mathbf{R}^{(m)}=\tilde{\mathbf{R}}^{(m-1)}+\frac{t-1}{t+2}\left(\tilde{\mathbf{R}}^{(m)}-\tilde{\mathbf{R}}^{(m-1)}\right)$
		\IF{$\ell_{\omega}(\mathbf{R}^{(m)})< \ell_{\omega}(\mathbf{R}^{(m-1)})$}
		\STATE~~~~$t = t +1$
		\ELSE
		\STATE
		~~~~$t=1$
		\ENDIF
		\IF{$\ell_{\omega}(\mathbf{R}^{(m)})\leq\varepsilon$}
		\STATE ~~~~$\hat{\mathbf{R}}=\mathbf{R}^{(m)}$
		\STATE
		\textbf{Break}
		\ENDIF
	\end{algorithmic}
\end{algorithm}

\subsection*{Initialization}
As is stated in the main text, the optimization problem \eqref{optimization} is non-convex and its numerical method may be sensitive to the initialization. 

\begin{figure}[htb]
	\includegraphics[width=1\textwidth]{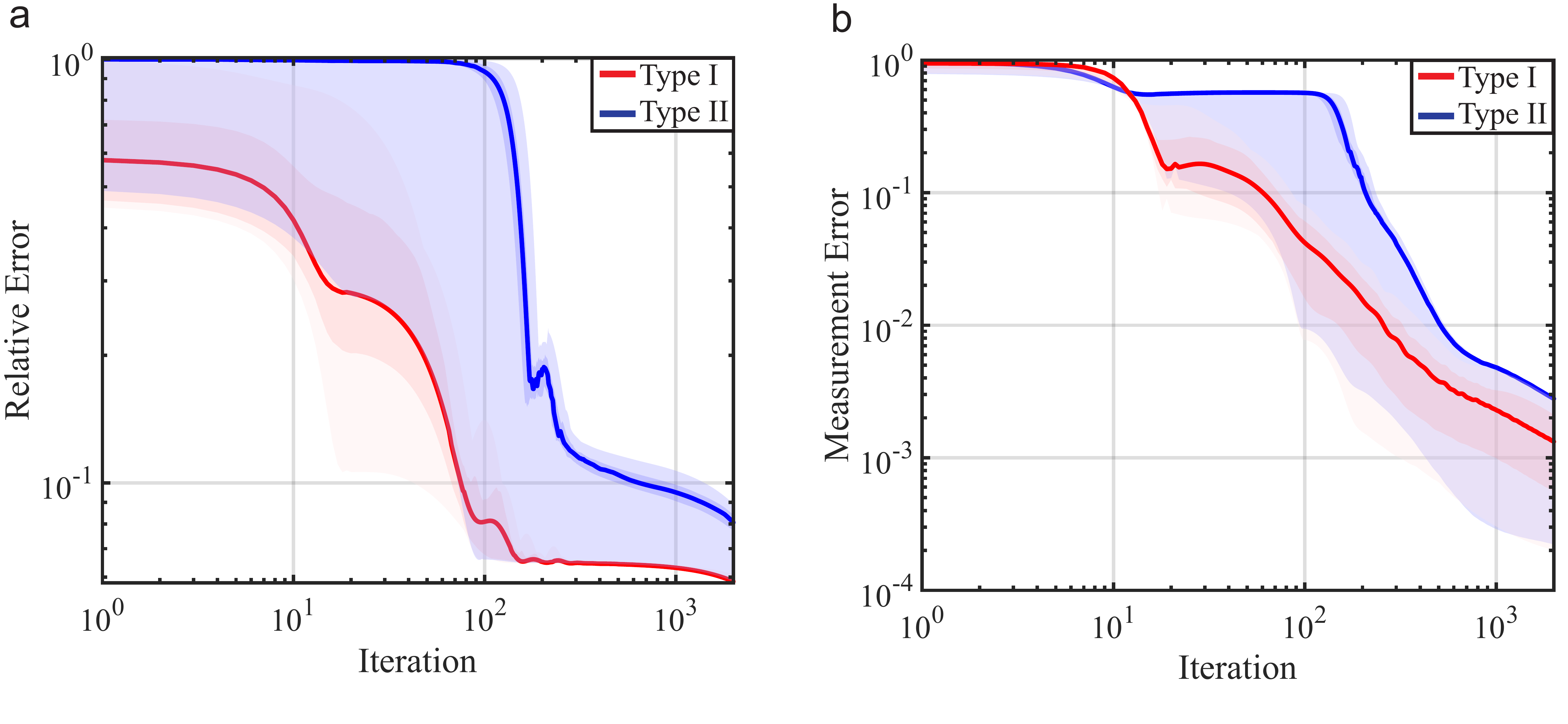}
	\caption{\textbf{a} and \textbf{b} record the relative error and measurement error recovered by the NPRS method at each iteration with two different types of initialization. The thick line shows the median over 25 trials, light area is between min and max, darker area indicates 25th and 75th quantile.} 
	\label{comparison}
\end{figure}

In this section, simulations will be applied to test the performance of NPRS method by using two different types of initialization. The type I is $\alpha\bm{1}$ where $\alpha$ is randomly chosen from $(0,1]$ as the initialization and $\bm{1}$ is a length $n$ vector with each entry equal to $1$. Each entry of the type II initialization is sampled from the Gaussian distribution $\mathcal{N}(0,0.1)$.  We choose the numerical results presented in Fig.~2\textbf{b} as the signal of interest. The range of time is set from 3ns to 165ns. $\Delta_{\text{step}}=1\Gamma_0$, $t_{\text{step}}=0.1974$ ns, $\bm{\Omega}=-150\Gamma_0:150\Gamma_0$, and $\mathbf{O}=-100\Gamma_0:100\Gamma_0$, where `:' means the range with interval 1$\Gamma_0$. The total iteration is 2000. For each initialization, we randomly generate 25 samples and record their measurement errors and relative errors respectively. The definition of relative error is shown below:
\begin{eqnarray}
\sqrt{\frac{\sum_{\Delta\in\mathbf{O}}	\left|\hat{\mathbf{R}}(\Delta)-\mathbf{R}(\Delta)\right|^2}{\sum_{\Delta\in\mathbf{O}}\left|\mathbf{R}(\Delta)\right|^2}},
\label{tt22}
\end{eqnarray}
At the same time, the measurement error is formulated as 
\begin{eqnarray*}
	\sqrt{\frac{\sum_{t_k\in\mathbf{E}}\sum_{\Delta_D^l\in\bm{\Omega}}	\left(I(k,l)-\left|\mathbf{F}_{:,k}^{\text{H}}(\hat{\mathbf{R}}\odot\mathbf{T}_{l})\right|^{2}\right)^2}{\sum_{t_k\in\mathbf{E}}\sum_{\Delta_D^l\in\bm{\Omega}}I^2(t_k,\Delta_D^l)}}.
	\label{tt21}
\end{eqnarray*}
The results are shown in Fig.~\ref{comparison}. We can find that the NPRS method is robust to both initializations and converges quickly even with the random initialization. At the same time, results from Fig.~\ref{comparison} also show that the type I initialization can make NPRS method converge faster. So, in the following tests, only type I initialization is used in the NPRS method.

Compared with the vanilla gradient descent algorithm mentioned in (4) in the main text, NPRS method introduces two modules below to make the gradient descent algorithm more robust and converging more quickly. 

\subsection*{Weighted loss}
Based on \eqref{optimization}, we introduce its weighed variant to prevent overfitting:
\begin{small}
	\begin{eqnarray}\label{newoptimization}
	\mathop{\textrm{minimize}}_{\mathbf{R}\in\mathbb{C}^n}\ell_{\omega}(\mathbf{R})=\sum_{k=1}^{K}\sum_{l=1}^{L}\omega(k,l)\left(\left|\mathbf{F}_{:,k}^{\text{H}}(\mathbf{R}\odot\mathbf{T}_{l})\right|^{2}-I(k,l)\text{log}\left(\left|\mathbf{F}_{:,k}^{\text{H}}(\mathbf{R}\odot\mathbf{T}_{l})\right|^{2}\right)\right),
	\end{eqnarray}  
\end{small}
where $\omega(k,l)=\omega_1(k,l)\omega_2(l)$ is the non-negative weight function to take different effects on the loss depending on $(k,l)$. The ideas behind the formulation of $\omega(k,l)$ contain two folds. 

First, $\omega_2(l)$ is used to balance the loss between different $l$, which is chosen as the Gaussian function
$$\omega_2(l)=e^{-\frac{(\Delta_D^l-u)^{2}}{2 \sigma^{2}}}$$ with
$u = \frac{1}{|\bm{\Omega}|}\sum_{\Delta_D^l\in\bm{\Omega}}\Delta_D^l$ which is in the neighbor of $0$ usually, and $\sigma^2=\frac{1}{|\bm{\Omega}|}\sum_{\Delta_D^l\in\bm{\Omega}}(\Delta_D^l-u)^2$. 
In many tests we find that   $\sum_{k=1}^{K}(I(k,l)-|\mathbf{F}_{:,k}^{\text{H}}(\mathbf{R}\odot\mathbf{T}_{l})|^{2})^2$ has a large difference for $l$ during the iteration. Concretely, the error above is much less when $|\Delta_D^l|$ is large comparing to which when $|\Delta_D^l|$ is small. Thus, to prevent over-fit so that $\left|\mathbf{F}_{:,k}^{\text{H}}(\mathbf{R}\odot\mathbf{T}_{l})\right|^{2}$ can fit $I(k,l)$ better when $|\Delta_D^l|$ is small, we use the Gaussian $\omega_2(l)$ as the weight function.

Second, $\omega_1(k,l)$ is formulated as below:
\begin{equation}\label{omega1}
\omega_1(k,l)=\left\{
\begin{aligned}
0~~~~~~~~~~~&,\text{if}~g(k,l)\leq p(k,l)\\
\frac{\left|\mathbf{F}_{:,k}^{\text{H}}(\mathbf{R}\odot\mathbf{T}_l)\right|}{\sqrt{I(k,l)}}~~&,\text{otherwise}
\end{aligned}
\right.,
\end{equation}
where  
$g\left(k,l\right):=
\left|\mathbf{F}_{:,k}^{\text{H}}(\mathbf{R}\odot\mathbf{T}_{l})\right|^{2}-I(k,l)+I(k,l)\text{log}(\frac{I(k,l)}{\left|\mathbf{F}_{:,k}^{\text{H}}(\mathbf{R}\odot\mathbf{T}_{l})\right|^{2}})$, which is non-negative. $g(k,l)=0$ when $\mathbf{R}$ is the ground truth, and $p(k,l)$ is a given parameter which can be adjusted according to different situations. The idea behinds \eqref{omega1} can be divided into two folds. First, since what we actually measured by the APDs are the counts of photons which are integers, the error caused by the truncation must be considered. As a result, we require $\left|\mathbf{F}_{:,k}^{\text{H}}(\mathbf{R}\odot\mathbf{T}_{l})\right|^{2}$ in the $p(k,l)$ neighbor of $I(k,l)$ to overcome over-fit. At the same time, the value of $p(k,l)$ also depends on $I(k,l)$. Specifically, $p(k,l)$ may be larger if $I(k,l)$ has more counts where more tolerances will be considered. In numerical simulations, $p(k,l)$ can approximate to $0$ because of the optimal simulation condition. Second, when $g\left(k,l\right)>p(k,l)$, let $ \omega_1(k,l)=\frac{\left|\mathbf{F}_{:,k}^{\text{H}}(\mathbf{R}\odot\mathbf{T}_l)\right|}{\sqrt{I(k,l)}}.$
As Ref.\cite{Wang2016Solving} stated,  $\omega_1(k,l)$ can be viewed as the confidence score on the reliability or meaningfulness of the corresponding gradient, so we here use it as a factor of weight function to measure the importance of the loss terms. 

Not only the adaptive weights but also the acceleration module will be combined to further improve the performance of NPRS method.

\subsection*{Acceleration and restart}
To accelerate the convergence of the numerical algorithm, we use  momentum acceleration \cite{1983A} algorithm at each iteration, namely,
\begin{equation}
\begin{array}{l}

\tilde{\mathbf{R}}^{(m)}=\mathbf{R}^{(m-1)}-\mu \nabla \ell_{\omega}\left(\mathbf{R}^{(m-1)}\right)\\
\mathbf{R}^{(m)}=\tilde{\mathbf{R}}^{(m)}+\frac{t-1}{t+2}\left(\tilde{\mathbf{R}}^{(m)}-\tilde{\mathbf{R}}^{(m-1)}\right).
\end{array}
\end{equation}
By acceleration, the gradient method may converge as fast as $\mathcal{O}(1/m^2)$ in optimal. At the same time, because the problem is also non-convex, we use the restart technique to make the acceleration more robust and prevent the algorithm to get stagnated easily. Specifically, if $\ell_{\omega}(\mathbf{R}^{(m)})$ does not decrease sufficiently, namely, $\ell_{\omega}(\mathbf{R}^{(m)})\geq \ell_{\omega}(\mathbf{R}^{(m-1)})$, it indicates that the result of the current iteration isn't satisfactory. So, it had better diminish the influence caused by the momentum in the next iteration and let $t=1$. 
\begin{figure}
	\includegraphics[width=0.9\textwidth]{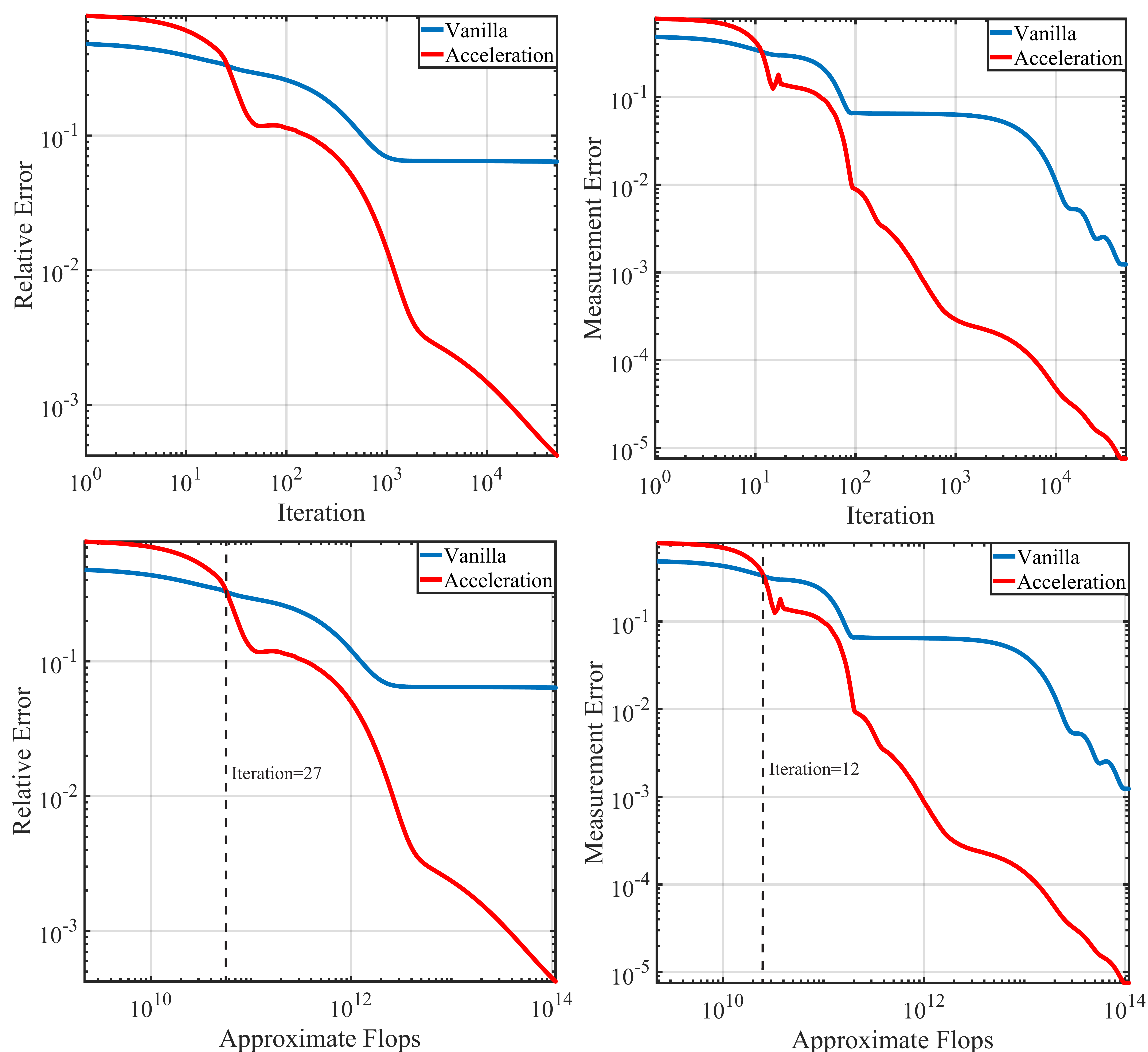}
	\caption{\textbf{a} and \textbf{b} record the relative error and measurement error recovered by the NPRS method with vanilla gradient and acceleration module at each iteration. \textbf{c} and \textbf{d} depict the relationship between the computational cost of different methods and the relative error or the measurement error, respectively.}
	\label{different method}
\end{figure} 
\subsection*{Computational cost of NPRS method}
The computational complexity of the NPRS method is dominated by the matrix production in calculating the gradient and the weights. Other steps only add a negligible amount of computation cost. The details are shown in TABLE \ref{ct}. So the total computational cost of each step is approximated $5n+3nL+2KLn+20KL+3Ln\text{log}n$. 

\begin{table}
	\caption{Baseline flop counts of NPRS at each iteration. Flop counts of $O(1)$ and pre-computations have been omitted.}
	\label{ct}
	\begin{tabular}{|c|c|}
		\hline Per iteration computation & Flops \\ 
		\hline$\nabla \ell_{\omega}\left(\mathbf{R}^{(m-1)}\right)$ & $2Ln\text{log}n+2KLn+2nL+12KL$ \\
		\hline $\mathbf{R}^{(m)}$ via formulation & $5n$ \\
		\hline 	update auxiliary $t$& $nL+Ln\text{log}n+8KL$ \\
		\hline 
	\end{tabular}
\end{table}

Next, simulations will be implemented to illustrate the performance of NPRS method using acceleration module compared with vanilla gradient. The total iterations are both $50000$. The common initialization is $\bm{1}.$ Other parameters are the same with the simulations in Fig.~\ref{comparison}. Similarly, the computational cost at each iteration for NPRS with vanilla gradient is $2n+2nL+2KLn+4KL+2Ln\text{log}n$. The results are shown in Fig.~\ref{different method}. We can find that the NPRS method combined with the acceleration modules can improve the speed of convergence. At the same time, although the acceleration module will introduce extra computational cost,  we can find that it actually demands fewer total computations to achieve a lower error from Fig.~\ref{different method}(\textbf{c-d}).
\subsection*{Numerical simulations to prove the conjecture}
In this subsection,  numerical simulations will be applied to prove the conjecture that $\mathbf{R}$ is unique in the region $\mathbf{O}$. We also choose the numerical results presented in Fig~.2\textbf{b} as the signal of interest. The set of Doppler detuning has three kinds of types namely $\bm{\Omega}=-50\Gamma_0:50\Gamma_0$, $\bm{\Omega}=-100\Gamma_0:100\Gamma_0$, and $\bm{\Omega}=-150\Gamma_0:150\Gamma_0$. Other parameters are the same with the simulations in Fig.~\ref{comparison}. We generate $\mathbf{I}$ according to different $\bm{\Omega}$, and use NPRS method to recover $\mathbf{R}$ from the corresponding $\mathbf{I}$ respectively. The results are shown in Fig.~\ref{system}. 

From Fig.~\ref{system}(\textbf{a-d,f,g,i,j}), we can find that $\hat{\mathbf{R}}$ recovered by NPRS fits the ground truth $\mathbf{R}$ well in the range of $\bm{\Omega}$. But the transmission of $\hat{\mathbf{R}}$ cannot fit $\mathbf{R}$ well in those regions out of $\bm{\Omega}$. At the same time, the measurement error at each case is around $10^{-5}$. 
This can fully verify the conjecture. As a result, in order to recover $\mathbf{R}$ well within the the region $\mathbf{O}$, the range of the Doppler detuning set $\bm{\Omega}$ must cover $\mathbf{O}$. 
\begin{figure}[!htb]
	\includegraphics[width=0.95\textwidth]{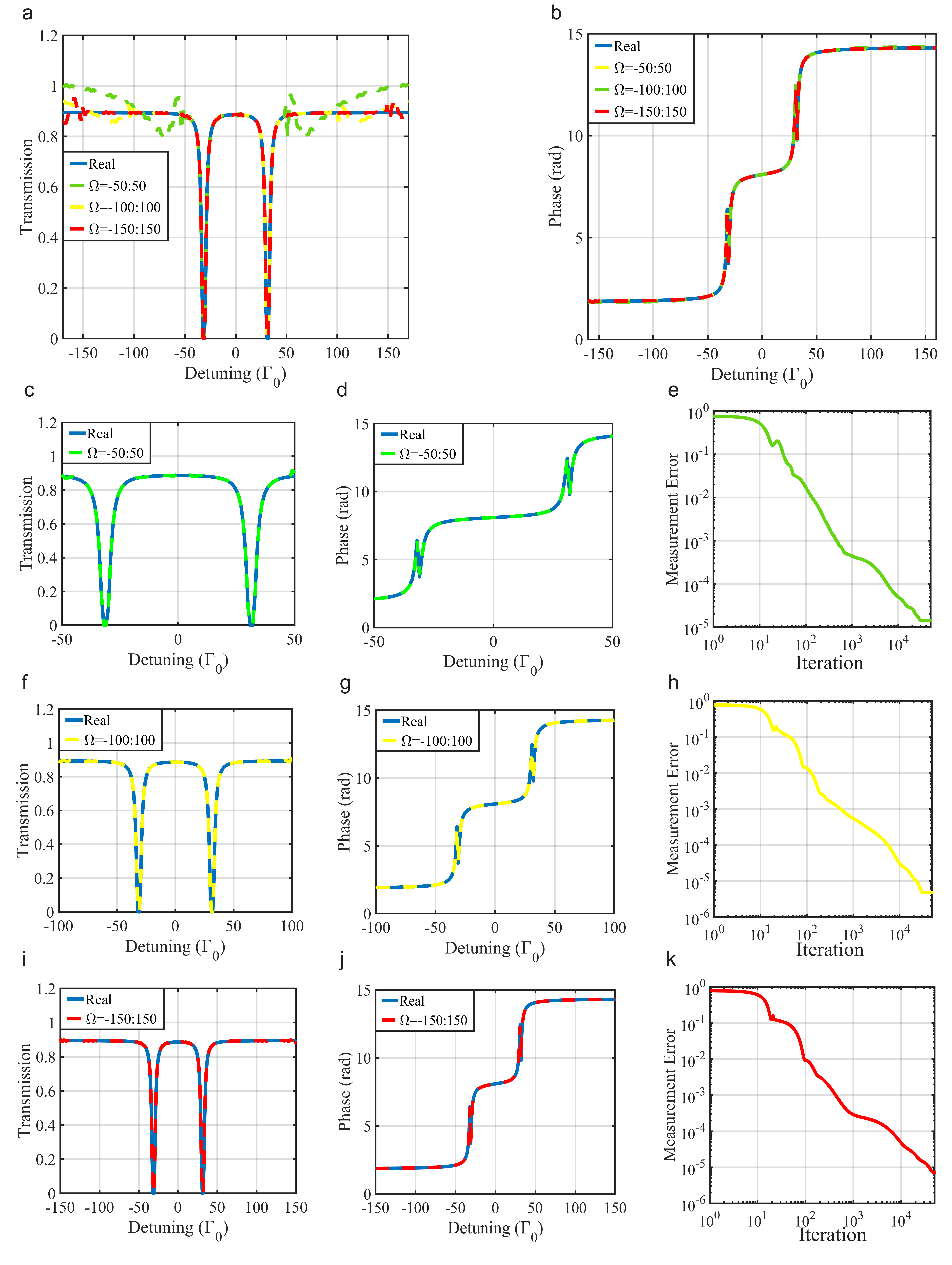}
	\caption{\textbf{a},\textbf{b} the spectrum and phase recovered by the NPRS method under three kinds of $\bm{\Omega}.$ When $\bm{\Omega}=-50\Gamma_0:50\Gamma_0$, \textbf{c-d} is the spectrum and phase recovered by the NPRS method, \textbf{e} is the measurement error at each iteration.  \textbf{f-h} and \textbf{i-k} are the results when $\bm{\Omega}=-100\Gamma_0:100\Gamma_0$ and $\bm{\Omega}=-150\Gamma_0:150\Gamma_0$.}
	\label{system}
\end{figure} 
\end{document}